\documentclass[
aps,
pra,
superscriptaddress,
preprint,
12pt,
onecolumn,
titlepage,
floatfix
]
{revtex4-2}
\usepackage{graphicx}
\usepackage[colorlinks,
            linkcolor=blue,
            citecolor=blue,
            anchorcolor=blue,
            urlcolor=blue,
            ]{hyperref}
\usepackage{leftidx}
\usepackage{subfigure}
\usepackage{amsmath}
\usepackage{amssymb}
\usepackage{epstopdf}
\usepackage{amsmath}
\usepackage{lineno}

\newlength{\onecolfig}
\newlength{\twocolfig}
\newcommand{\figref}[1]{\mbox{Fig.~\ref{#1}}}
\newcommand{\figpanel}[2]{Fig.~\hyperref[#1]{\ref*{#1}(#2)}}
\newcommand{\figpanels}[3]{Fig.~\hyperref[#1]{\ref*{#1}(#2)-(#3)}}
\newcommand{\figpanelNoPrefix}[2]{\hyperref[#1]{\ref*{#1}(#2)}}

\newcommand{\unit}[1]{\,\mbox{#1}}

\newcommand{\MHz}{\unit{MHz}}
\newcommand{\GHz}{\unit{GHz}}

\newcommand{\um}{\unit{$\mu$m}}
\newcommand{\nm}{\unit{nm}}

\newcommand{\us}{\unit{$\mu$s}}

\begin{document}
\title{Minimizing Kinetic Inductance in Tantalum-Based Superconducting Coplanar Waveguide Resonators for Alleviating Frequency Fluctuation Issues}
\author{Dengfeng Li}
\email{dengfengli@tencent.com}
\affiliation{Tencent Quantum Laboratory, Tencent, Shenzhen, Guangdong 518057, China}
\author{Jingjing Hu}
\affiliation{Tencent Quantum Laboratory, Tencent, Shenzhen, Guangdong 518057, China}
\author{Yuan Li}
\affiliation{Tencent Quantum Laboratory, Tencent, Shenzhen, Guangdong 518057, China}
\author{Shuoming An}
\email{shuomingan@tencent.com}
\affiliation{Tencent Quantum Laboratory, Tencent, Shenzhen, Guangdong 518057, China}

\date{\today}
\begin{abstract}
Advancements in the fabrication of superconducting quantum devices have highlighted tantalum as a promising material, owing to its low surface oxidation microwave loss at low temperatures. 
However, tantalum films exhibit significantly larger kinetic inductances compared to materials such as aluminum or niobium.
Given the inevitable variations in film thickness, this increased kinetic inductance leads to considerable, uncontrolled frequency variances and shifts in components like superconducting coplanar waveguide resonators.
Achieving high precision in resonator frequencies is crucial, particularly when multiple resonators share a common Purcell filter with limited bandwidth in superconducting quantum information processors. 
Here, we tackle this challenge from both fabrication and design perspectives, achieving a reduction in resonator frequency fluctuation by 100 fold. 
Concurrently, the internal quality factor of the superconducting coplanar waveguide resonator remains at high level. 
Our findings open up new avenues for the enhanced utilization of tantalum in large-scale superconducting chips.
\end{abstract}
\maketitle
\section{Introduction}
Over the past two decades, significant advancements in device coherence times have propelled superconducting qubits to the forefront of quantum computing technology~\cite{de2021materials, murray2021material}. 
The lifespan of planar transmon qubits, fabricated using tantalum (Ta), has been extended over 0.5ms~\cite{place2021new,wang2022towards,bal2023systematic}.
Historically, research of superconducting materials has primarily focused on minimizing uncontrolled defect sites, which could potentially function as sources of two-level systems (TLS)~\cite{simmonds2004decoherence,muller2019towards}. 
However, the kinetic inductance, another significant property of superconducting material, has received less attention in the context of superconducting quantum information processors.

In comparison to conventional superconducting materials like aluminum or niobium (Nb), Ta displays a higher kinetic inductance, depending on factors such as film thickness $d$ and temperature~\cite{gubin2005dependence,yu2005fabrication}. 
The increased kinetic inductance, together with the thickness variation $\Delta d$ of magnetron-sputtered body-centered cubic or $\alpha$-Ta films, may lead to considerable uncontrolled superconducting coplanar waveguide (SCPW) resonator frequency fluctuations across the wafer~\cite{watanabe1994kinetic,gao2008physics}. 
Such fluctuations present a significant challenge for large-scale circuit QED-based quantum information systems, as they can cause frequency crowding and cross-talk issues, particularly when multiple resonators share a limited-bandwidth common Purcell filter~\cite{blais2021circuit,li2023optimizing}.

Firstly, we will investigate the impact of kinetic inductance on the SCPW resonator frequency~\cite{doyle2008lumped, gao2008physics}. 
In this study, a quarter-wavelength SCPW resonator is utilized, and its frequency is estimated as: $f = \frac{1}{4l\sqrt{CL}}$, where $l$ denotes the SCPW length, and $C$ ($L$) represents the capacitance (inductance) per unit length. 
The magnetic field penetrates the superconductor to a depth governed by the penetration depth $\lambda$~\cite{kittel2018introduction}, as shown in \figpanel{fig:R}{b}.
For an intuitive understanding, we can image that the supercurrent within this layer under the driving of a alternating filed will be accelerated and gain velocity and momentum. 
When the field is reversed, copper pairs in the supercurrent must first loose their momentum before changing direction. 
The current lags the field by 90$^\circ$ due to the kinetic energy stored within the copper pairs. 
That is inductance.
Consequently, $L = L_{\rm m}+L_{\rm k}$, where $L_{\rm m}$ is the constant geometric component, corresponding to the magnetic energy outside the superconducting film, and $L_{\rm k}$ is the variable kinetic component. 
Of course, there is also magnetic energy inside the superconducting film, but we find it will not change our optimization strategy, we will not consider it in this experimental study. 
Because there is an inevitable thickness variation $\Delta d$ when sputtering the Ta film depends on deposition system. 
Consequently, our main goal is to minimize $L_{\rm k}$, expecting a decrease in its variance $\Delta L_{\rm k}$ and resonator frequency fluctuations with the same $\Delta d$ of the Ta film.

Next, we will clarify the elements influencing $L_{\rm k}$, and the methods to minimize it. 
As shown in Ref.~\cite{watanabe1994kinetic}, if $d$ is smaller than twice the magnetic penetration depth $\lambda$, the kinetic inductance can be approximated as:
\begin{equation}
L_{\rm k} = \mu_{0}\frac{\lambda^2}{d w}g(s,w,d),
\label{Lk}
\end{equation}
where $\mu_{0}$ denotes the vacuum permeability, $w$ the SCPW's center electrode width, $s$ the gap between the center and ground electrodes, as shown in \figpanel{fig:R}{b}, and $g(s,w,d)$ a geometric factor. 
Although the assumption $d<2\lambda$ for this approximation may not always be valid in this study, the formula still provide an intuitive understanding of how reducing $\lambda$ and optimizing the SCPW geometry can contribute to the minimization of $L_{\rm k}$.

To decrease $\lambda$, we can enhance the film thickness $d$, as shown in studies on Nb and aluminum films~\cite{gubin2005dependence,lopez2023magnetic}. 
Also, increasing the substrate temperature during sputtering, which will lead to larger $\alpha$-Ta grain sizes~\cite{hylton1989effect, yu2005fabrication, jones2023grain}, can also reduce $\lambda$.
This is based on the idea that a larger grain size allows fewer disturbances in the superconducting current flow, behaving more like a bulk material, which usually has a smaller $\lambda$ than a thin film. 
Besides fabrication optimization, $L_{\rm k}$ can also be reduced by adjusting $s$ and $w$ from a design perspective. 
By applying these modifications, we achieve a more than 100-fold reduction in resonator frequency variation.
 
\section{Results}
\subsection{Device and Optimized Results}
\begin{figure*}[!htb] 
\begin{center} 
\includegraphics[width=1\textwidth]{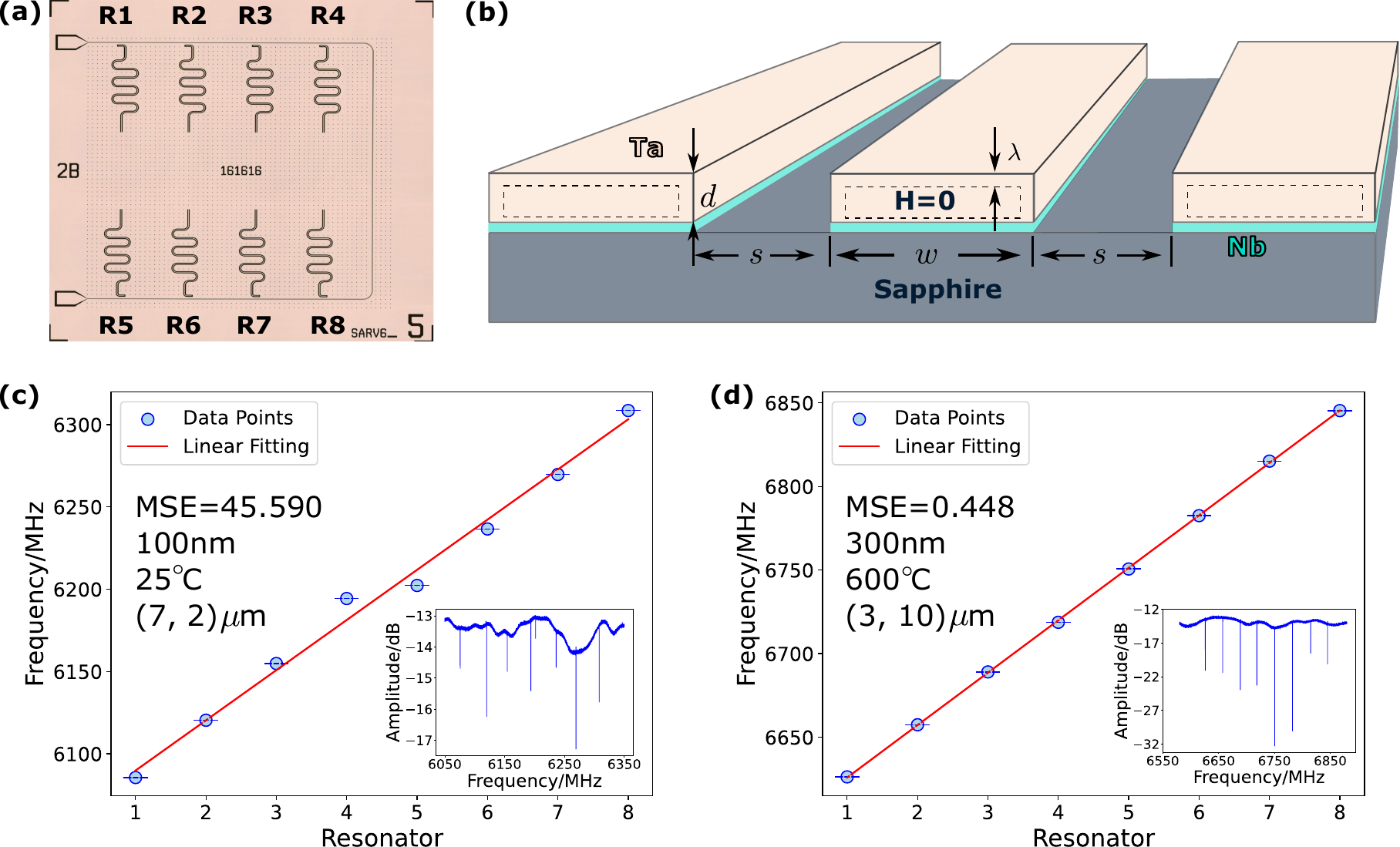} 
\end{center} 
\caption{
\textbf{Overview of the system under investigation and the optimization results.}
(a) Optical micrographs of an 8 mm$\times$8 mm resonator test chip featuring eight coupled resonators (R1-R8) with a shared readout line. 
The average designed frequency is 6.7 GHz with a 30 MHz frequency gap. 
(b) SCPW schematic, where a tantalum (Ta) film is sputtered on a sapphire substrate with a niobium buffer layer. 
In cryogenic measurements, a penetration depth $\lambda$ exists for the Ta film with thickness $d$.
(c) Pre-optimization resonator frequencies, showing measured frequencies (blue points) and linear fittings (red lines). 
Error bars indicate ±1 standard deviation of the fitted frequencies, which is too small to reveal more than the bar hat. 
The subplot shows the amplitude response of the transmissive coefficient $S_{21}$. 
The Ta film has a thickness of 100 nm, and the sapphire substrate temperature is maintained at room temperature (around 25$^\circ$C) during sputtering. 
For this chip, the values of ($s, w$) are (7, 2)\um, resulting in an MSE of 45.59.
(d) Post-optimization resonator frequencies where the Ta film thickness is 300 nm, and the sapphire substrate temperature is 600$^\circ$C. 
Here, the ($s, w$) values are (3, 10)\um. 
The measurement condition stays consistent with resonators in (c). 
With these optimization methods, we get an MSE of 0.448 - a decrease of over 100 fold compared to the original value.
The mean frequency has also increased by 539 MHz following the optimization.
}
\label{fig:R} 
\end{figure*}
We begin by introducing the SCPW resonator design under investigation. 
As depicted in \figpanel{fig:R}{a}, eight resonators are coupled to a shared readout line. 
We add labels only in the figure for identification. 
The resonators' frequencies are designed to follow a linear distribution, with an average frequency of 6.7\GHz~and a 30\MHz~frequency gap between each. 
The coupling quality factor $Q_{\rm c}$ of the resonators falls within the range of 0.5 to 1 million. 
All electromagnetic simulations are conducted with a frequency convergence precision of one percent, without considering the kinetic inductance. 
The relative permittivity of the sapphire substrate is set to $\epsilon_{\rm r}=$10.55~\cite{krupka1999complex}.
The internal quality factor $Q_{\rm i}$ and frequency of each resonator can be obtained by fitting their transmissive signal $S_{21}$, measured from the readout line~\cite{mcrae2020materials}.

To quantify the resonator frequency variation, we opt for the mean square error (MSE) of the linear fit to the observed frequency distribution, rather than a traditional parameter like the standard deviation of a group of resonators with the same design frequency. 
MSE is defined as MSE=$\frac{1}{n}\sum_{i=1}^n(y_i-\hat{y}_i)^2$, where $n$ represents the number of data points, $y_i$ refers to the actual values, and $\hat{y}_i$ denotes the predicted values. 
This approach is chosen due to the requirement for distinct frequencies among different resonators on the same readout line. 
As a result, we use MSE as an indicator of deviation from linearity, where a smaller value is desirable.
Additionally, we compute the mean frequency shift relative to the design value 6.7\GHz~to represent the relative amount of $L_{\rm k}$ for various optimization parameters.

In \figpanel{fig:R}{b}, we show the cross section of the SCPW resonator.  
In our study, we also adjust the substrate temperature during sputtering to modify the grain size of the $\alpha$-Ta film. 
However, to ensure a high purity $\alpha$-Ta, previous studies~\cite{place2021new, wang2022towards} have indicated that a high temperature for sapphire substrate is necessary. 
To address this challenge, we employ a 5\nm~Nb buffer layer to get high purity $\alpha$ phase of Ta for different temperatures~\cite{alegria2023two}, as illustrated in \figpanel{fig:R}{b}. 

The pre-optimization results are displayed in \figpanel{fig:R}{c}. 
During this chip's fabrication, we maintained the sapphire substrate temperature at room temperature, approximately 25$^\circ$C. 
The Ta film thickness was set to 100\nm, and the SCPW resonator structure had an $s$ of 7\um~and a $w$ of 2\um. 
Blue points represent measured SCPW resonator frequencies, while error bars indicate $\pm$ one standard deviation of the fitting process of the transmissive coefficient $S_{21}$. 
The subplot demonstrates the amplitude response of $S_{21}$. From these measurements, we obtained an MSE of 45.590.

\figpanel{fig:R}{d} presents SCPW resonator frequencies after optimization. 
We applied the following optimization methods: increasing the Ta film thickness to 300\nm, setting the sapphire substrate temperature to 600$^\circ$C during Ta film sputtering, and adjusting $s$ and $w$ to 3\um~and 10\um~for the SCPW resonator. 
All SCPW resonators are measured with the same condition with SCPW resonators in \figpanel{fig:R}{c}.
These optimizations yielded an MSE of 0.448, signifying a more than 100-fold reduction in resonator frequency variation compared to the pre-optimization MSE. 
Additionally, utilizing these optimization methods, the discrepancy between the average measured frequencies of the resonators and the average frequencies of the designed resonators has been reduced from approximately -500 MHz to approximately +36 MHz. 
Despite the positive sign being attributed to the uncertainty of the sapphire permittivity during the SCPW resonator design process, this result signifies a fundamental reduction of $L_{\rm k}$.

\subsection{Film Thickness Optimization}
\begin{figure*}[!htb] 
\begin{center} 
\includegraphics[width=1\textwidth]{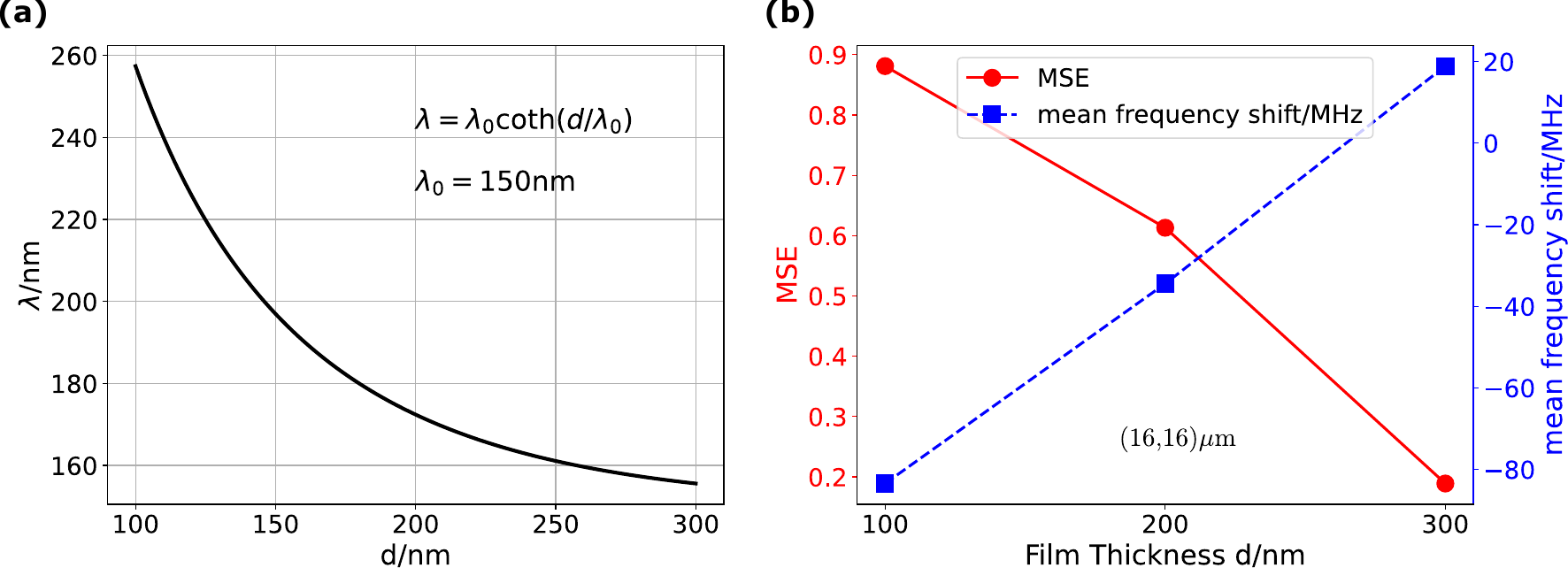} 
\end{center} 
\caption{
\textbf{Influence of Film Thickness on Resonator Frequencies.}
(a) Theoretical effective penetration depth $\lambda$ as a function of film thickness $d$ and bulk penetration depth $\lambda_0$.
(b) Mean Squared Error (MSE) and the mean frequency shift relative to the design mean value of the resonator groups for varying film thickness $d$.
The film is sputtered at room temperature (25$^\circ$C) on a sapphire substrate, with both slot and center widths of the SCPW resonators set at 16\um.
Due to the uncertainty of the permittivity in simulation, the mean frequency shift is positive for $d=300$ nm.
However, this does not affect the conclusion that the thickest film exhibits minimized $L_{\rm k}$ and the smallest resonator frequency variance.
} 
\label{fig:d} 
\end{figure*}
For the Ta film, the theoretical effective penetration depth can be represented as~\cite{gubin2005dependence, lopez2023magnetic}:
\begin{equation}
\lambda=\lambda_0\coth(d/\lambda_0),
\label{lambda}
\end{equation}
where $\lambda_0$=150nm is the penetration depth of bulk superconductor~\cite{greytak1964penetration} and $d$ is the thickness of the film. 
The trend of the effective penetration depth changes in Ta films is illustrated in \figpanel{fig:d}{a}, where an increase in film thickness results in a decrease in effective penetration depth. 
As can be inferred from Eqn.~\ref{Lk}, the increase in $d$ and the corresponding decrease in penetration depth contribute to a reduction in kinetic inductance. 
This is attributed to the fact that $g(s,w,d)$ changes very slowly with $d$. 
Consequently, kinetic inductance can be mitigated by increasing the film thickness.

\figpanel{fig:d}{b} displays the MSE distribution for different samples with only film thickness as varying parameters. 
As the thickness of Ta films increases from 100 nm to 300 nm, the MSE decreases monotonically from about 0.9 to approximately 0.2, signifying a notable improvement in resonator frequency stability. 
Furthermore, we employ $\Delta f$ of SCPW resonators to examine the changes in kinetic inductance. 
Since $f\propto \frac{1}{\sqrt{L}}$, the change in $\Delta f$ from approximately -80\MHz~to +20\MHz, relative to the designed frequency of 6.7\GHz, corresponds to a decreased $L_{\rm k}$ as the Ta film thickness increases. 
The positive $\Delta f$ stems from the uncertainty in the permittivity of sapphire, which is set to 10.55 in the simulation and may be slightly higher than the actual value of the material used.

\subsection{Sputtering Temperature Optimization}
\begin{figure*}[!htb] 
\begin{center} 
\includegraphics[width=1\textwidth]{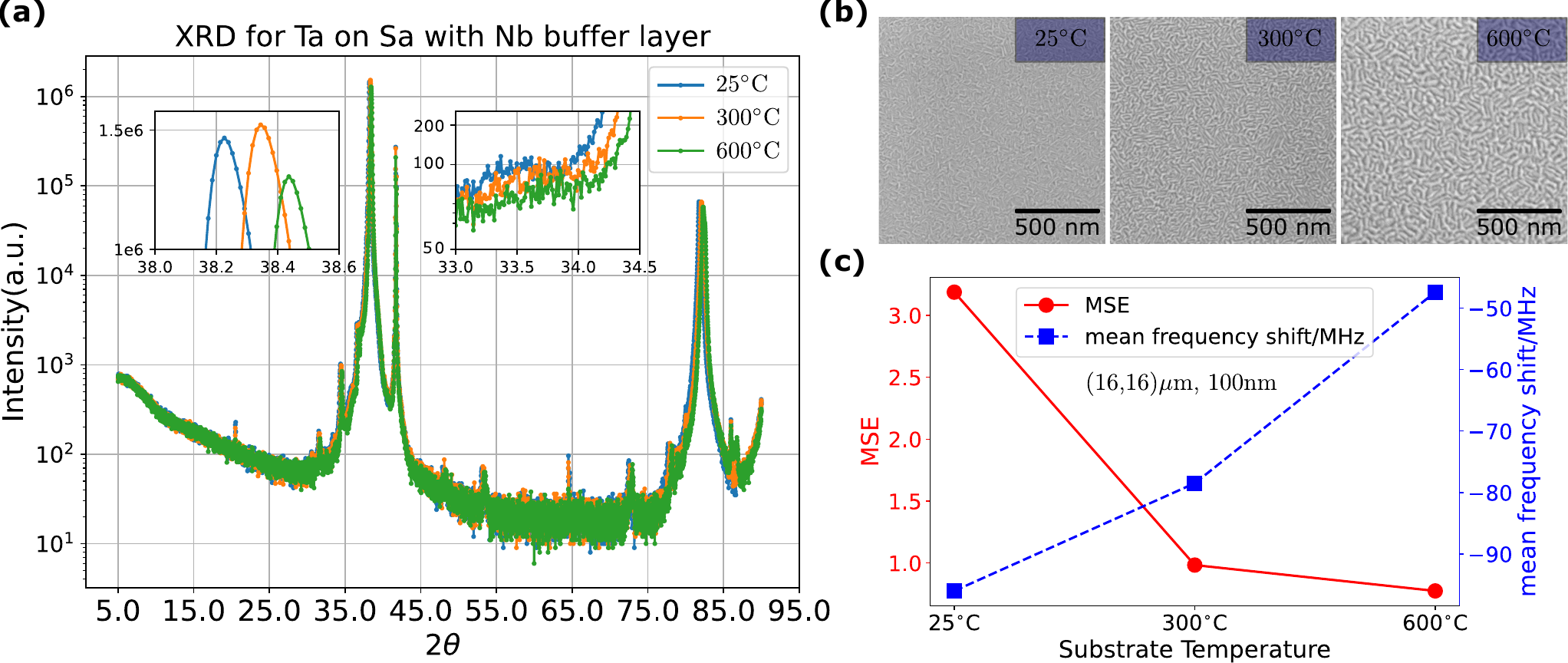} 
\end{center} 
\caption{\textbf{Effect of Substrate Temperature during Sputtering Process.}
(a) X-ray Diffraction (XRD) analysis of the Ta film deposited on a Sapphire substrate with a Nb buffer layer.
The two insets display the nearly consistent $\alpha$-Ta peak and the absence of $\beta$-Ta for different temperatures.
(b) Comparison of grain sizes using Scanning Electron Microscopy (SEM) images of the film prepared at various temperatures.
(c) Experimental MSE and $\Delta f$ of the resonators fabricated at different temperatures.
As the temperature increases to 600$^\circ$C, MSE decreases by more than three times. 
Additionally, $\Delta f$ is reduced by approximately 50 MHz.
} 
\label{fig:t} 
\end{figure*}
In addition to Ta film thickness, the substrate temperature during Ta sputtering is another crucial factor affecting $\lambda$ and $L_{\rm k}$. 
In general, higher temperatures lead to larger grain sizes and consequently fewer grain boundaries~\cite{jones2023grain}, resulting in a structure that more closely resembles the bulk superconductor, characterized by a smaller $\lambda_0$.
However, altering the temperature of the sapphire substrate changes the lattice matching conditions between sapphire and Ta, leading to different Ta film phases and other phases may have larger loss than $\alpha$~\cite{shiojiri2003preparation, wu2023high, place2021new, wang2022towards}. 
To maintain the purity of the $\alpha$-phase, we introduce a 5 nm Nb buffer layer to ensure good lattice matching across different temperatures~\cite{alegria2023two}. 
As shown in \figpanel{fig:t}{a}, we employ X-ray diffraction (XRD) to analyze the purity of $\alpha$-Ta. 
In the upper right inset, the primary $\alpha$ peak remains nearly constant at various temperatures. 
The slight differences in diffraction arise from minor distortions in the $\alpha$-Ta lattice~\cite{colin2017origin}.
In the upper middle inset, we also do not observe $\beta$-Ta, which locates at 33.7$^\circ$.
In \figpanel{fig:t}{b}, we utilize scanning electron microscopy (SEM) to confirm the grain size change with different substrate temperatures.
It is evident that a higher temperature leads to larger grain sizes.

We fabricate the test chip using the same design, only varying the substrate temperature during sputtering. 
The low-temperature results are summarized in \figpanel{fig:t}{c}. 
As the temperature increases to 600$^\circ$C, MSE decreases by more than three times. 
Additionally, $\Delta f$ is reduced by approximately 50\MHz. 
These findings suggest that a higher temperature will both reduce $L_{\rm k}$ and enhance the resonator frequency stability against film thickness variations.

\subsection{Geometry Optimization}
\begin{figure*}[!htb] 
\begin{center} 
\includegraphics[width=1\textwidth]{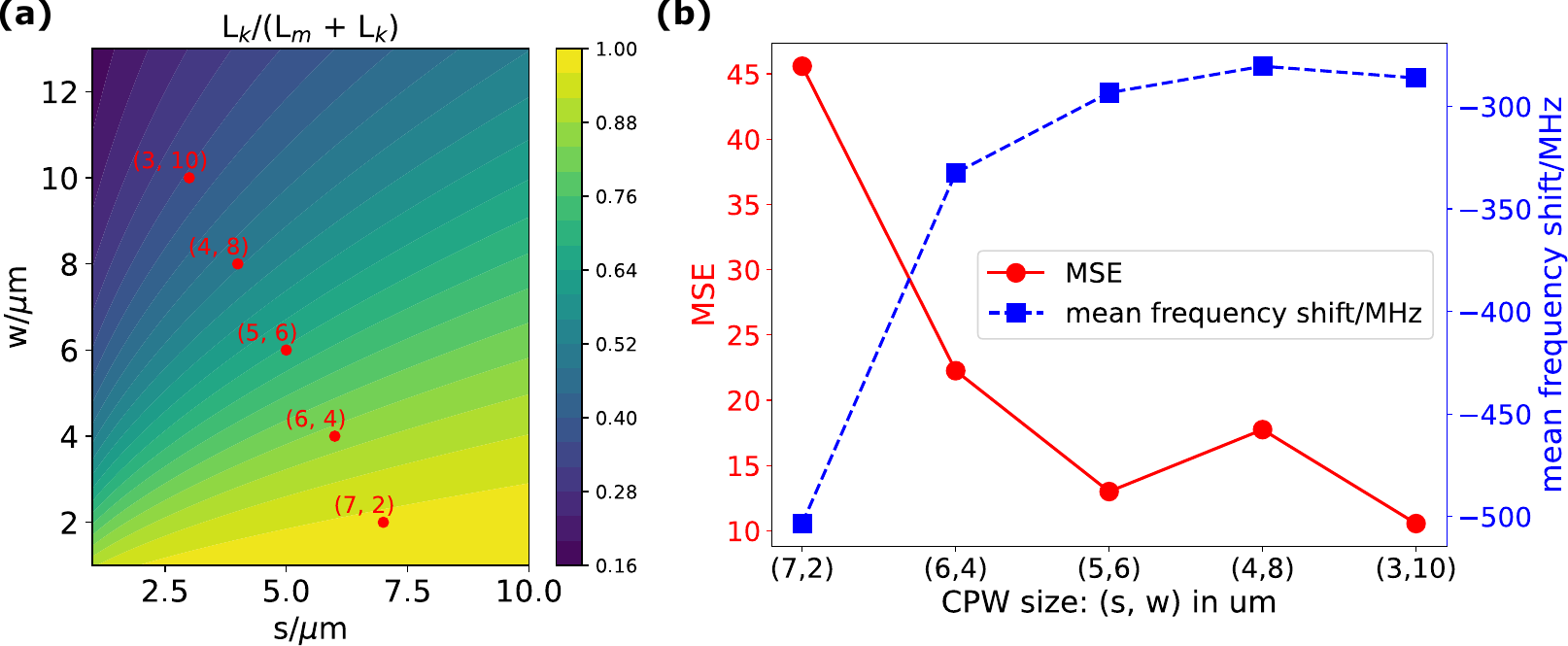} 
\end{center} 
\caption{
\textbf{Influence of SCPW Slot Gap and Center Conductor Dimensions.}
(a) Theoretical ratio of kinetic inductance to total inductance, assuming that the film thickness d is less than twice the bulk penetration depth $\lambda$ and that the current density is uniformly distributed across the film thickness. 
The formula used to generate this figure is taken from Ref.~\cite{watanabe1994kinetic}. 
However, as per Ref.~\cite{gao2008physics}, this formula should only be used to provide an intuitive guidance for the optimization direction.
(b) Experimental MSE and $\Delta f$ of the resonators designed with different ($s, w$), where $s$ represents the slot gap size and $w$ denotes the center conductor size, respectively.
These chips are fabricated at room temperature with a 100 nm thickness of $\alpha$-Ta.
} 
\label{fig:sw} 
\end{figure*}
In prior studies, the width of the center electrode and the gap between the center electrode and the ground electrode were kept at 16$\mu$m. 
However, such wide SCPW resonators may not fit into the limited space available on large-scale superconducting quantum processors. 
Therefore, it is essential to optimize a set of SCPW resonator dimensions ($s,\ w$) to achieve low kinetic inductance within a fixed space constraint, such as $w+2s= 16$\um.

We calculate the kinetic inductance ratio to the total inductance $L_{\rm k}/(L_{\rm m}+L_{\rm k})$ using Eqn.~\ref{Lk} and $L_{\rm m}$ in Ref.~\cite{watanabe1994kinetic, gao2006experimental,gao2008physics} with varying electrode width $w$ and the gap between the center electrode and ground electrode $s$, as shown in \figpanel{fig:sw}{a}.
It is clear qualitatively that smaller $s$ and larger $w$ result in a smaller kinetic inductance ratio, which is preferred.
We should keep in mind that the quantitative values in \figpanel{fig:sw}{a} may be inaccurate, because the formula we use is valid only when $d\ll 2\lambda$, but this may not hold true in our chip.

We maintain the total size of the SCPW resonator at 16\um~and choose a series of ($s,\ w$) values.
We ensure the mean resonator frequencies are always 6.7\GHz.
The low-temperature results of MSE and $\Delta f$ are displayed in \figpanel{fig:sw}{b}.
As $s$ decreases from 7\um~to 3\um, the MSE drops from about 45 to approximately 10, and the absolute $\Delta f$ decreases by 200\MHz.
These findings indicate that a smaller gap and larger center width can effectively reduce $L_{\rm k}$ and improve the resonator frequency stability against film thickness variations. 
Moreover, it is also suggested that when designing Purcell filters and resonators, it is preferable to maintain the same ($s,\ w$) values to minimize frequency mismatching caused by the difference in $L_{\rm k}$.

\subsection{High Internal Quality Factor of the Optimized Resonator}
\begin{figure*}[!htb] 
\begin{center} 
\includegraphics[width=0.5\textwidth]{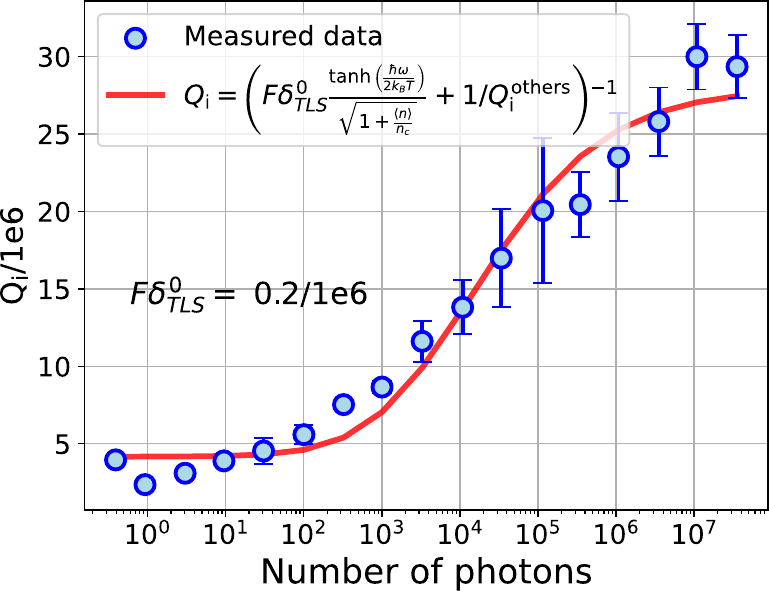} 
\end{center} 
\caption{
\textbf{Measurements of the resonator's internal quality factor as a function of photon flux number for the final optimized fabrication parameters.} 
Specifically, the thickness of the Ta film is approximately 300\nm, and the sapphire substrate temperature during Ta film sputtering is 500$^\circ$C. 
Dots represent the experimental measurements, while the red line indicates the best fit to the TLS model. 
Error bars denote ±1 standard deviation during the fitting of the resonator's $Q_{\rm i}$.
} 
\label{fig:q} 
\end{figure*}
In superconducting quantum computing applications, we utilize a (16,16)\um~SCPW resonator with a 300 nm thickness, sputtered at 500$^\circ$C. 
The internal quality factor, $Q_{\rm i}$, of this resonator has an empirical relationship with the qubit relaxation time ($T_1$) at 5\GHz. 
In our routine tests, we have observed that when the $Q_{\rm i}$ of this resonator (6.7\GHz) reaches $x\times 10^4$, the highest $T_1$ of a single qubit fabricated with the same film can achieve $x$\us~(5\GHz) in our lab.
The $Q_{\rm i}$ of the SCPW resonator is measured at different microwave power levels, as shown in \figref{fig:q}. 
Error bars represent ±1 standard deviation while fitting the $Q_{\rm i}$ of the resonator.
The microwave flux photon number is calculated from the incident microwave power, the fitted coupling quality factor (about 0.7 million), the resonator frequency (6636 MHz), and the calculated impedance of the readout waveguide and the resonator~\cite{mcrae2020materials}.
At single photon power, the $Q_{\rm i}$ is approximately 2.5 million, which is similar to the best reported results on sapphire without a Nb buffer layer~\cite{crowley2023disentangling}. 
With this fabrication technique, according to the experience in our lab, we predict a best $T_1$ of qubit to be 250\us~at 5\GHz. 
We fit the $Q_{\rm i}$ of the resonator measurements using the formula:
\begin{equation}
Q_{\rm i}=\left(F \delta_{\rm T L S}^0 \frac{\tanh \left(\frac{\hbar \omega}{2 k_{\rm B} T}\right)}{\sqrt{1+\frac{\langle n\rangle}{n_{\rm c}}}}+1/Q_{\rm i}^{\rm others}\right)^{-1},
\end{equation}
resulting in a filling factor-loss angle product $F \delta_{\rm T L S}^0=0.2/1\text{e}6$, which is 
one of the lowest result for the 16\um~gap~\cite{mcrae2020materials}, providing evidence for the suitability of our technology for high-coherence superconducting quantum chips.

\section{Discussion}
Our study introduces a comprehensive method to minimize kinetic inductance in Ta-based SCPW resonators, effectively addressing frequency crowding issues. 
By carefully controlling Ta film thickness and substrate temperature during sputtering, and optimizing SCPW geometry, we achieved a significant reduction in uncontrolled frequency shifts and a 100-fold reduction in frequency variance characterized by MSE, crucial for the precision needed in multi-qubit systems. 
The high internal quality factor of 2.5 million at the single-photon power level, combined with the reduced kinetic inductance, highlights the potential of our approach for high-coherence, large-scale superconducting quantum chips.

These findings not only present a comprehensive strategy for optimizing Ta-based superconducting circuits but also provide insights into the impact of microstructure on superconducting properties, thereby laying the foundation for further enhancements in quantum device performance. 
By reversing the kinetic inductance reduction optimization methods detailed here, our work can be directly applied to nonlinear quantum devices that require large kinetic inductance, such as microwave kinetic inductance detectors~\cite{day2003broadband} and kinetic inductance traveling wave parametric amplifiers~\cite{klimovich2022traveling}.

\section{Method}
\subsection{Fabrication}
To maintain high-purity $\alpha$-Ta under various sputtering conditions, especially at different temperatures, all Ta films are fabricated with an Nb buffer layer. 
First, a thin Nb buffer layer is deposited on a sapphire substrate using dc magnetron sputtering at room temperature. 
Second, Ta films with the desired thickness and temperature conditions are deposited using the same sputtering system. 
Next, we use ultraviolet exposure to define the resonators on the chip using a photoresist. 
After a brief development, an inductively coupled plasma (ICP) system is employed to remove unwanted Ta films. 
The wafer then undergoes ultrasonication with acetone and isopropanol for five minutes each. 
Additionally, the wafer is cleaned in a piranha solution to eliminate residual resist. 
Finally, the sample is diced into chips, and the protective resist is removed using a remover at 80$^\circ$C for over 12 hours.

\subsection{Low Temperature Measurement}
All resonator $S_{21}$ measurements are conducted at approximately 13 mK in the mixing chamber (MXC) of a cryostat. 
The input attenuation for the chip includes 20 dB on the 3K layer, 3 dB on the still layer, 6 dB on the cold plate, and 40 dB on the MXC~\cite{krinner2019engineering}. 
The output signal passes through two double junction isolators, a 40 dB gain High Electron Mobility Transistor (HEMT) at the 4K layer, and a final 40 dB gain room temperature amplifier. 
The $S_{21}$ measurements are performed using a Vector Network Analyzer (VNA) with a bandwidth precision of up to 1 Hz, and the bandwidth and repetitions are adjusted based on the criterion that the relative deviation of $Q_{\rm i}$ should be less than 20\%. 
Two infrared (IR) filters are connected to both the input and output ports of the chip, which is packaged in a 4-port aluminum box and thermally attached to the mixing chamber.

The rescaled $S_{21}$ is fitted using the formula~\cite{khalil2012analysis}: $S_{21}(f) = 1 - \frac{Q / \hat{Q}_c}{1 + 2 i Q \frac{f - f_0}{f_0}}$, where $\hat{Q}_c = Q_c e^{-i \phi}$ represents the complex coupled quality factor.
The photon number reaching the resonator is calculated based on the power arriving at the chip and the external coupling quality factor $Q_{\rm c}$, which is also determined from the fitting of experimental data~\cite{mcrae2020materials}.
To calculate the power reaching the chip, the input power from the VNA, in addition to the attenuators mentioned earlier, is reduced by an additional 10 dB to account for cable loss, which is independently calibrated. 

\bibliography{MRF}

\section{Data and Code availability}
All data needed to evaluate the conclusions in the paper are present in the paper or the supplementary materials.
Codes used in the theoretical calculation are available from the corresponding author on reasonable request.
\section{Author contributions}
S.M.A. and D.F.L. planned the experiment and designed the chip. 
D.F.L., J.J.H., and Y.L. fabricated the device and performed room temperature measurements.
D.F.L. and S.M.A. conducted low temperature measurements and analyzed the data. 
All authors contributed to the manuscript, which was composed by S.M.A. and D.F.L.
\section{Competing Interests Statement.}
The authors declare no competing interests.
\end{document}